\documentclass[aps,pra,twocolumn,superscriptaddress]{revtex4-1}

\usepackage{graphicx}
\usepackage{color}
\usepackage{hyperref}
\usepackage{amssymb}

\begin{document}


\title{Single atom counting in a two-color magneto-optical trap}


\author{Martin Schlederer}
\email[Corresponding author: ]{mschlede@physnet.uni-hamburg.de}
\affiliation{Institut f\"{u}r Laserphysik, Universit\"{a}t Hamburg, 22761 Hamburg, Germany}
\affiliation{The Hamburg Centre for Ultrafast Imaging, Universit\"{a}t Hamburg, 22761 Hamburg, Germany}

\author{Alexandra Mozdzen}
\affiliation{Institut f\"{u}r Laserphysik, Universit\"{a}t Hamburg, 22761 Hamburg, Germany}
\affiliation{The Hamburg Centre for Ultrafast Imaging, Universit\"{a}t Hamburg, 22761 Hamburg, Germany}

\author{Thomas Lompe}
\affiliation{Institut f\"{u}r Laserphysik, Universit\"{a}t Hamburg, 22761 Hamburg, Germany}
\affiliation{The Hamburg Centre for Ultrafast Imaging, Universit\"{a}t Hamburg, 22761 Hamburg, Germany}

\author{Henning Moritz}
\affiliation{Institut f\"{u}r Laserphysik, Universit\"{a}t Hamburg, 22761 Hamburg, Germany}
\affiliation{The Hamburg Centre for Ultrafast Imaging, Universit\"{a}t Hamburg, 22761 Hamburg, Germany}

\date{\today}

\begin{abstract}
Recording the fluorescence of a magneto-optical trap (MOT) is a standard tool for measuring atom numbers in experiments with ultracold atoms.
When trapping few atoms in a small MOT, the emitted fluorescence increases with the atom number in discrete steps, which allows to measure the atom number with single-particle resolution.
Achieving such single particle resolution requires stringent minimization of stray light from the MOT beams, which is very difficult to achieve in experimental setups that require in-vacuum components close to the atoms.
Here, we present a modified scheme that addresses this issue: Instead of collecting the fluorescence on the MOT (D2) transition, we scatter light on an additional probing (D1) transition and collect this fluorescence with a high-resolution microscope while filtering out the intense MOT light.
Using this scheme, we are able to reliably distinguish up to 17 $^{40}$K atoms with an average classification fidelity of 95 \%. 

\end{abstract}


\maketitle

\section{Introduction\label{section:introduction}}
In recent years ultracold atoms have emerged as a unique and powerful platform to experimentally study few-body physics.
They provide synthetic model systems where key properties such as the particle number, the strength of the interparticle interactions, and the confining potential can be tuned nearly at will \cite{Jochim2020Higgs}. 
This has enabled pioneering works that study e.g. entanglement generation \cite{pezze2018metrologyNonclassicalReview} or the emergence of correlations and quantum phases with increasing particle number \cite{Jochim2013FermiSea,Jochim2020Higgs}. 

An essential requirement for such experiments is the ability to count the number of atoms in the trap with high fidelity \cite{Ott2016singleAtomdetectionReview}.
Collecting fluorescence light while laser cooling and trapping the atoms in a magneto-optical trap (MOT) has proven to be a reliable and well established method to measure atom numbers with single particle resolution \cite{Kimble1994,Meschede1996,Grangier2001sub,Jochim2011deterministic,Oberthaler2013accurate,Oberthaler2015double,Klempt2019preparation}.
As the number of photons emitted by a single atom is quite limited, these experiments require considerable effort to maximize the signal-to-noise ratio of the fluorescence signal. 
Common techniques for achieving this include the use of large magnetic field gradients to compress the MOT and thereby spatially concentrate the fluorescence signal, using optics with high numerical aperture to maximize the number of collected photons, as well as employing highly sensitive and low noise cameras.
At the same time, detrimental stray light collection is suppressed as far as possible by employing laser cooling beams with small diameters, avoiding the presence of unwanted scattering surfaces close to atoms, and spatial filtering of the fluorescence signal.

Using these techniques, single atom atom resolution for atom numbers in excess of 300 has been achieved \cite{Oberthaler2013accurate,Oberthaler2015double}.
However, the stringent limits on the amount of stray light present in the system render this method unsuitable for many experimental setups that were not specifically designed for this purpose.
Suppressing stray light detection is particularly challenging in experimental setups that feature in-vacuum components such as electrodes, high-resolution optics or atom chips, as spatial filtering of stray light that originates close to the atoms is inefficient. 

In this work, we describe a technique that addresses this issue by using two different transitions.
In brief, we cool and trap fermionic $^{40}$K atoms using light on the D2 transition but collect fluorescence on the D1 transition.
The D2 MOT light propagates along all three spatial directions and hence generates a large amount of stray light which can be filtered out before the camera with dichroic mirrors.
Meanwhile, D1 fluorescence photons are generated by a D1 pumping beam whose size and beam path have been chosen specifically to avoid scattering surfaces and thereby minimize the amount of stray light that is generated.
This scheme allows us to perform single atom counting of up to 17 atoms with an average classification fidelity of 95 \% in an experimental setup featuring a high-resolution microscope objective placed inside the vacuum chamber very close to the atoms.  

This work is structured as follows: the experimental setup is described in section \ref{section:setup}, followed by a discussion of the image post-processing and atom number determination in section \ref{section:evaluation}.
The classification fidelity and the limiting factors of the presented scheme are discussed in section \ref{section:fidelity}, and section \ref{section:conclusion} summarizes the presented work.

\section{Setup\label{section:setup}}

\begin{figure}
\center
\includegraphics[width=8.6cm]{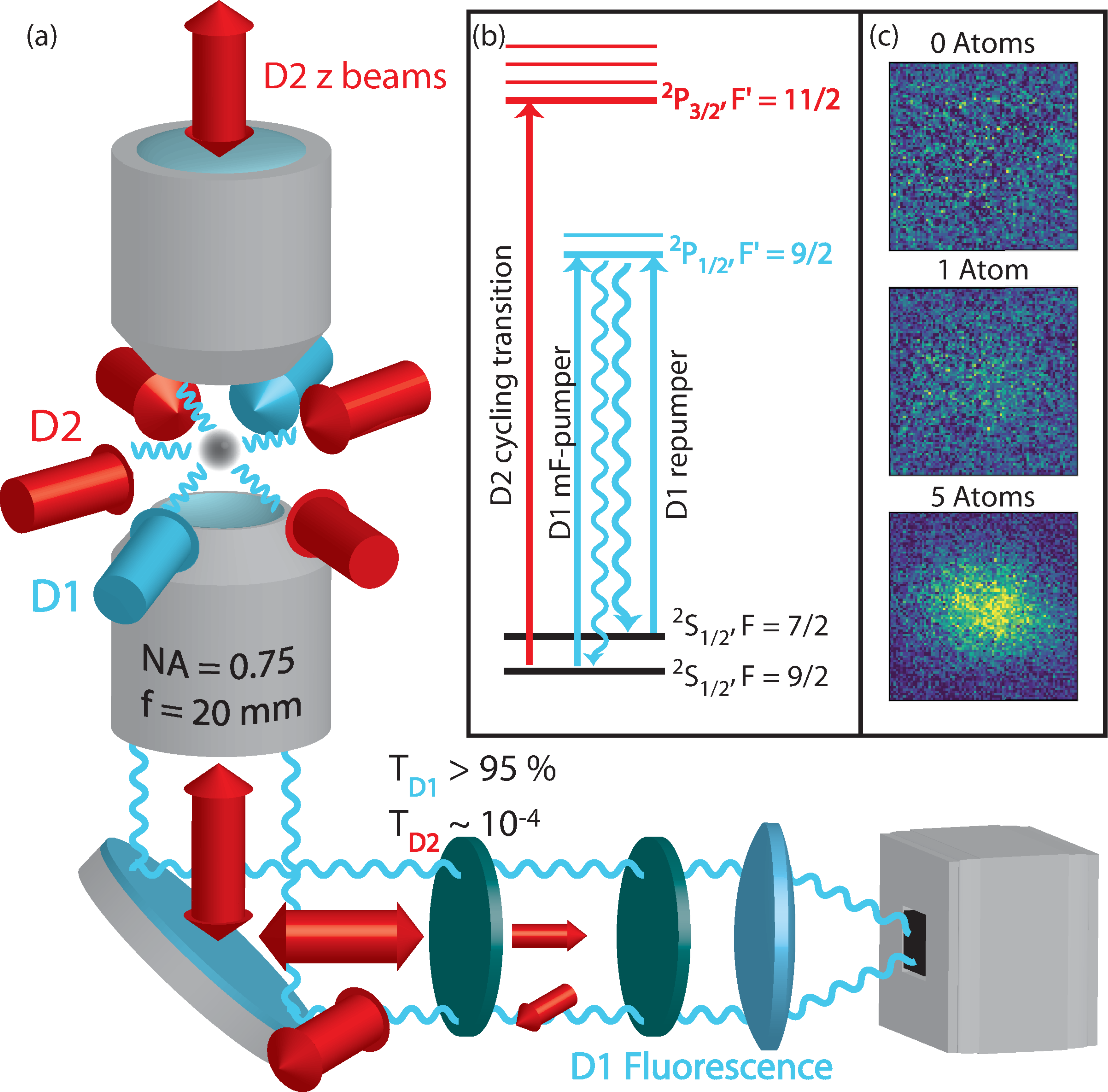}
\caption{\label{Fig1}
(a) Sketch of the experimental setup. 
The detection MOT is formed by three pairs of D2 trapping beams (dark red), while a pair of D1 beams (light blue) generates D1 fluorescence for detection.
The D1 fluorescence light is collected by an in-vacuum microscope objective (NA = 0.75, f=20\,mm), separated from the D2 trapping beams using dielectric filters, and finally imaged on an sCMOS camera.
(b) Simplified level scheme showing the atomic transitions utilized in the detection MOT: The D2 line is used for cooling and trapping the atoms, while the D1 repumper addresses atoms that decay into the F = 7/2 state. The D1 mF-pumper excites atoms from the ground state to the $^{2}$P$_{1/2}$ manifold and thereby enhances the D1 fluorescence.
(c) Typical fluorescence images containing 0, 1, and 5 atoms.
}
\end{figure}

The central element of our setup to study ultracold few-body systems is a pair of microscope objectives with high numerical aperture \footnote{Special Optics Type 54-31-20, N.A. = 0.75, working distance = 2.5 mm, effective focal length f = 20 mm, achromatic for 770 nm, 870 nm and 590 nm.} placed inside the vacuum chamber, see Fig. \ref{Fig1}(a).
These objectives are ideally suited to generate arbitrary potentials and image single atoms with high spatial resolution.
However, at the same time, they represent a significant challenge for performing atom counting with a MOT.

The first constraint placed by the objectives is that they cover such a large fraction of the solid angle that the vertical MOT beams have to pass through the objectives.  
This precludes using the objectives to detect the fluorescence created by the MOT light, as the fluorescence signal will be completely drowned out by the MOT beams.
At the same time the geometry of the vacuum chamber, which has to accommodate the in-vacuum components, severely limits the optical access for fluorescence light collection along other directions.
Combined with the large background of MOT light scattered by the objectives this makes it exceedingly difficult to measure the fluorescence on the MOT transition with a signal to noise ratio sufficient for single atom resolution.

Fortunately, both of these issues can be addressed by using our two-color approach.
By adding a retroreflected D1 pumping beam that propagates orthogonal to the objective axis, we create fluorescence light on the D1 transition while causing very little D1 stray light to enter the objective.
As the wavelengths of the D1 and D2 transition differ by $\sim$\,3nm, this allows us to collect the D1 fluorescence photons with one of the in-vacuo objectives while filtering out the D2 MOT light with two dichroic longpass filters \footnote{Alluxa LC-HBP770.7/3, nominal center wavelength 770.7 nm, FWHM: 3 nm, transmission $>$95 \% at 770.1 nm, blocking $>$OD4 at 766.7 nm.} with very steep edges transmitting in total more than 90 \% of the D1 light at 770.1\,nm but less than $10^{-8}$ of the D2 light at 766.7\,nm.

In the following, the D2 MOT and the D1 excitation scheme are described in more detail.
The D2 MOT is formed by three beam pairs addressing the cycling transition between the F = 9/2 and F’ = 11/2 hyperfine states. 
The horizontal D2 MOT beams have 1/$e^2$ beam waists of 1.5\,mm and are retroreflected, while the vertical beams are independent of each other and have waists of only 340 $\mu$m and 480 $\mu$m, as the short focal length of the microscope strongly constrains the maximal achievable beam size.
The magnetic quadrupole field for the MOT is provided by a set of magnetic field coils in an anti-Helmholtz configuration.
To decrease the influence of background light on the atom counting we use a strong quadrupole field with a gradient of $dB/dz$ = 132 G/cm in the vertical axis to make the MOT as small as possible and thereby concentrate the fluorescence signal in the smallest possible area.  
For our setup we find an optimum at a combined saturation parameter for the D2 light of 12 and 31 for the horizontal and vertical beams, respectively, and a detuning $\Delta \omega_{D2}$ of $-1.2\,\Gamma$, where $\Gamma/2\pi = 6$\,MHz is the natural linewidth of the excited state.
The detuning of the D2 MOT beams is a compromise between minimizing the MOT size, which is minimal for small detunings, and a stable scattering rate that is not too sensitive to detuning changes, which is the case at larger detunings. 
We note that our D2 MOT differs from a standard MOT in one aspect: There are no D2 beams that repump atoms that have decayed into the F = 7/2 hyperfine ground state and thereby became dark to the cycling transition.
Instead, this repumping is performed by one of the D1 excitation beams.

The D1 beams propagate in the horizontal plane along an axis having a $45^\circ$ angle to the D2 beams and have waists of 1.2\,mm.
The first beam addresses the F = 7/2 $\leftrightarrow$ F' = 9/2 transition. This repumper brings atoms that have decayed into the F = 7/2 hyperfine state back to the F = 9/2 ground state. It operates with a saturation parameter of 14 and a detuning $\Delta \omega_{D1}$ of +0.6 $\Gamma$.

However, using only this repumper does not produce enough D1 fluorescence for imaging as the F = 7/2 hyperfine state is populated only slowly while cooling on the D2 cycling transition \footnote{This is due to the fact that decay from the excited state F' = 11/2 of the D2 MOT cycling transition to the F = 7/2 state is forbidden and one has to rely on off-resonant excitations to the F' = 9/2 state in the D2 manifold, which is 8.5 $\Gamma$ detuned from the D2 laser frequency.}.
Hence, we use a second D1 beam to drive the F = 9/2 $\leftrightarrow$ F' = 9/2 transition. 
This so-called mF-pumper addresses the prevalent F = 9/2 ground state atoms and excites them to the F' = 9/2 manifold.
From there, the atoms decay into the F = 9/2 in about 30\% and the F = 7/2 manifold in about 70\% of the cases. 
If the atom is pumped into the F = 7/2 manifold, the subsequent repumping results in an average of three more photons being emitted on the D1 line before the atom is pumped back to the F = 9/2 manifold. 
By varying the strength of this mF-pumper we can therefore control the amount of D1 fluorescence emitted by the atoms. 
For our imaging the mF-pumper transition is driven resonantly with a saturation parameter of 7. 
The resulting significant amount of light pressure on the atoms is cancelled by retroreflecting the beam.

The D1 fluorescence light emitted by the atoms is collected by one of the microscope objectives and imaged onto an sCMOS camera, examples of typical fluorescence images obtained with the setup are shown in Fig. \ref{Fig1}(c).
The total collection efficiency of the imaging system is estimated to be 4.8$\pm$0.6\,\%, with the dominant effects being the relative solid angle covered by the microscope (17 \%) and the quantum efficiency of the camera (37 \% \footnote{Andor Zyla 5.5.}).
On average, the camera collects $6\cdot10^4$ photons per atom for an exposure time of 1\,s, corresponding to a scattering rate of 470\,kHz on the D1 transition.
Adding a retroreflected D1 beam pair along the second horizontal axis did not result in a significant improvement in the signal-to-noise ratio.
While the fluorescence increased by about 20\%, adding the additional beams also increased the amount of background light, resulting in a similar signal-to-noise ratio.

The number of background photons due to scattered light is $\sim$175/pixel, which for typical regions of interest of 120 x 120 pixels corresponding to 150 $\mu$m  x 150 $\mu$m amounts to about $2.5\cdot10^6$ photons. 
This background level is dominated by stray light from the D1 beams, the background originating from the D2 beam directly incident onto the objective accounts for less than 10\,\% of the background light.
To achieve this low background it is essential to not only perform the obvious filtering of D2 light with filters in front of the camera, but also to suppress the broad pedestal of incoherent light emitted by the semiconductor lasers and amplifiers generating the D2 beams. 
This is achieved by placing two ultra-narrow 780\,nm bandpass filters \footnote{Laseroptics Garbsen L-08729 IF780nm/6, FWHM $\approx$ 0.3 nm, transmission $\approx$ 90\%} angle-tuned to 766.7 nm in the beam path between the tapered amplifier and the fibers.

\section{Image post-processing and atom number extraction} \label{section:evaluation}

\begin{figure*}
	\center
	\includegraphics[width=17.8cm]{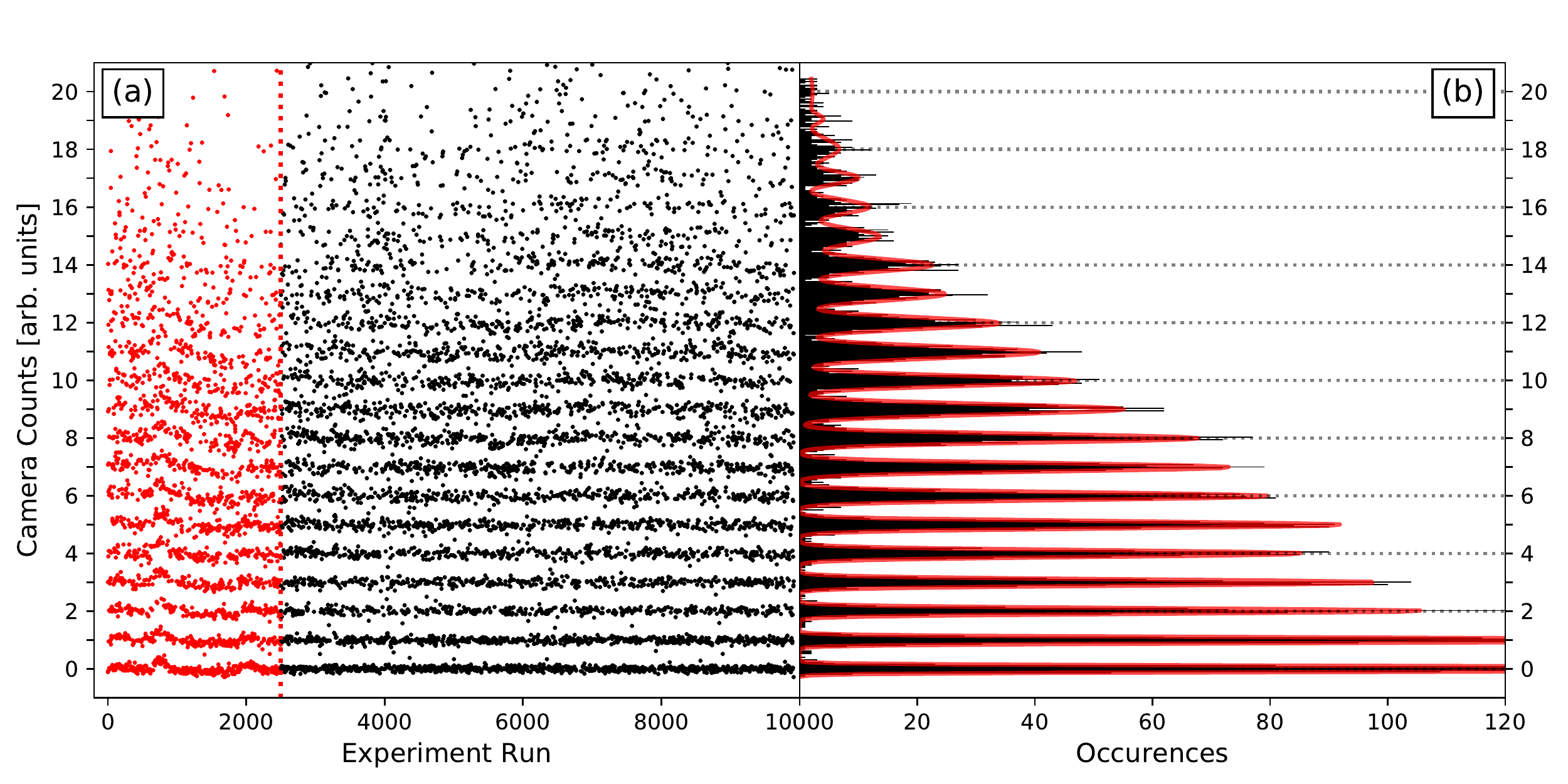}
	\caption{\label{Fig2} 
(a) Timeline of the fluorescence level over $\sim$10.000 experiment runs, where each dot shows the result of a single experiment run. 
The red dots to the left of the vertical dotted line show the raw fluorescence levels without correcting for time-dependent changes in offset and scattering rate, the black dots are corrected for these effects.
(b) Histogram of the corrected fluorescence levels.
The red line shows a fit to the data, consisting of a sum of 20 Gaussians with independent centers $x_n$, widths $\sigma_n$, and amplitudes.
We can reliably determine the atom number for systems with up to about 17 atoms.
Above this number, both the limited statistics of the data set and the width of the distributions prevent a reliable classification.
}
\end{figure*}
To characterize our imaging scheme and establish a protocol for determining the atom number from the fluorescence signal we generate a large number of sample images of MOTs containing a small number of atoms. 
We achieve this by releasing atoms from an optical trap and then recapturing them with the MOT.
This leads to a random number of atoms in the imaging MOT, where the mean value can be controlled via the atom number in the optical trap and the time delay between releasing the atoms and switching on the MOT.
While the two-color scheme described in section \ref{section:setup} yields fluorescence signals of single atoms that are significantly above the background noise (see Fig. \ref{Fig1}(c)), some steps of image post-processing are required to extract reliable particle numbers from this data.
In the first step, the background of scattered light is subtracted from the fluorescence images and the raw fluorescence signal is determined by summing up the camera counts within a defined region around the current center of the MOT.
Slow drifts in the fluorescence per atom and the zero atom offset are then determined and corrected by time-resolved fitting of these two parameters, thus providing a normalized fluorescence signal. 
A histogram of the normalized fluorescence signals, see Fig. 2(b), shows clearly discernable peaks corresponding to different atom numbers.
To assign atom numbers to individual images, we first quantify the positions and widths of the peaks by fitting a sum of Gaussians to the histogram.
We then define classification intervals around each peak based on their widths and assign atom numbers to all fluorescence measurements that fall within these intervals.

We now go through each step of the analysis in more detail, starting with the subtraction of the background signal.
To achieve this we make use of a principal component analysis, as this approach reduces noise compared to subtracting a reference image taken shortly after each atom image.
To this end, a set of reference images without atoms yet otherwise the same experimental conditions is taken and the main components are determined using singular value decomposition.
For the data shown in Fig. \ref{Fig2}, 100 reference images were taken at random times during the $\sim$10.000 runs and used for the singular value decomposition. The components with the ten highest singular values were then projected out of the actual data images \cite{Spreeuw2010fringeRemoval} by subtracting these basis vectors, weighted with the scalar product between them and the atom image.
Using more reference images or projecting out more components does not result in a significant improvement in the signal-to-noise ratio, as the components are not necessarily orthogonal to the fluorescence signal.

Next, the current MOT position on each of these background-free images is determined by fitting a gaussian to an averaged image obtained from the current and the ten preceding and following images.
This allows us to compensate for slow drifts of the MOT position over time and use a tight region of interest of only 150 $\mu$m x 150 $\mu$m for summing up camera counts. 
The resulting camera counts recorded by the experiment are shown as a function of time in Fig. \ref{Fig2}(a) (red dots).
While discrete steps in the fluorescence for different atom numbers can be clearly made out by eye, there are slow drifts in the signal levels over time and simple thresholds are not sufficient to reliably distinguish between different atom numbers with high fidelity.
To provide the time-resolved values of the zero atom offset and of the fluorescence per atom, we determine the current average count levels for 0, 1, 2, and 3 atoms from the 20 realizations closest in time containing these atom numbers, respectively, and fit a linear relation to these levels.
The camera counts are then corrected for these drifts, resulting in the black data points shown in Fig. \ref{Fig2}(a).

\begin{figure}
\center
	\includegraphics[width=8.6cm]{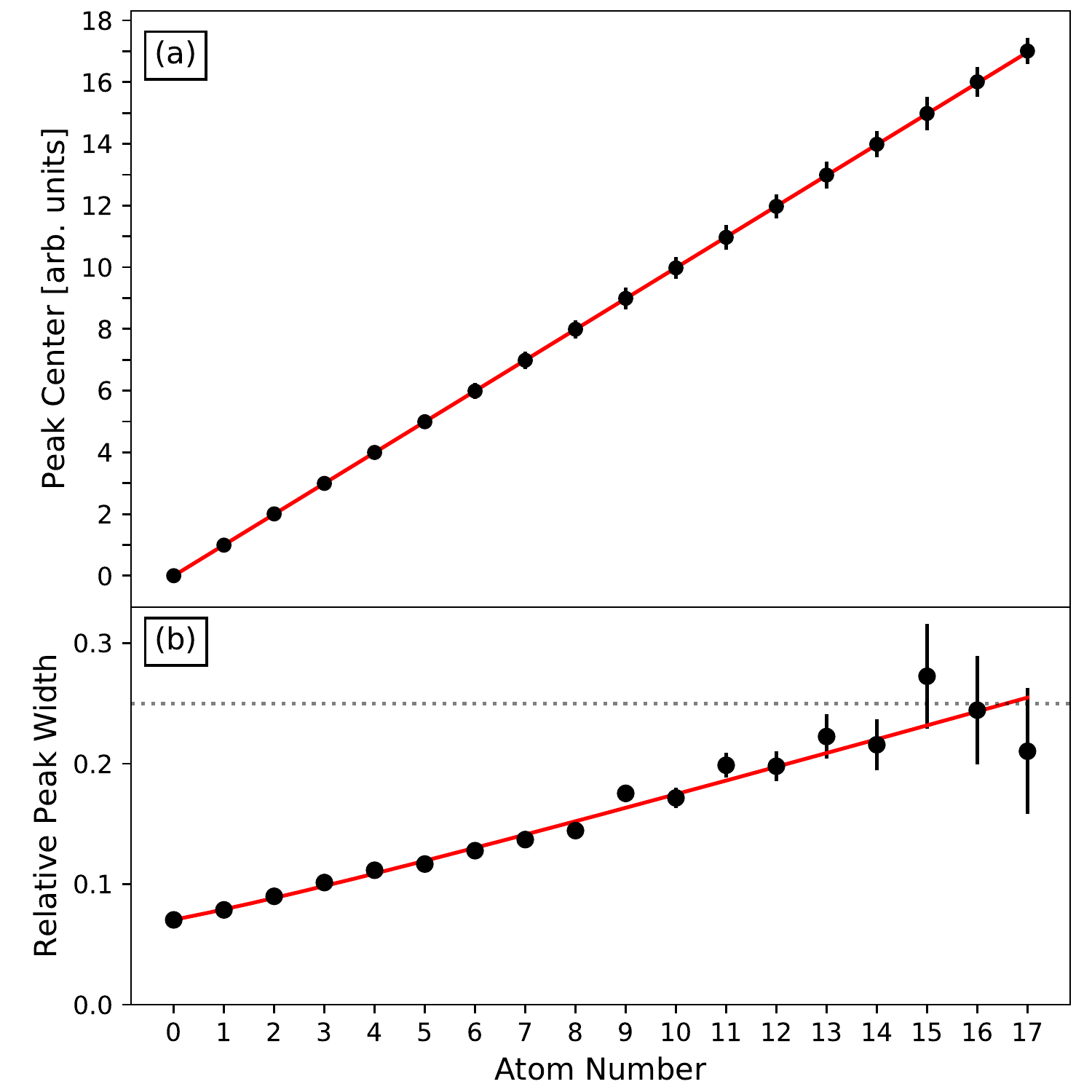}
	\caption{\label{Fig3} 
Mean value $x_n$ (a) and width $\sigma_n$ (b) of the fluorescence signal as a function of atom number, as determined from the fit to the histogram shown in Fig. \ref{Fig2}(b).
The mean fluorescence increases linearly with atom number (red solid line), indicating that density dependent effects are negligible. 
The error bars represent the $\pm2\sigma_n$ confidence intervals.
The widths $\sigma_n$ of the peaks in the histogram are normalized to the average fluorescence per atom and shown in panel (b).
We assume the peak widths to be a quadratic sum of a constant background noise, photon shot noise scaling with $\sqrt{N}$ and a linear noise contribution due to short-term variations in the intensity and detuning of the imaging light. 
Fitting the data with this model (red line) shows that the noise is dominated by the linear term.
The dotted line shows the point where the $\pm2\sigma_n$ classification regions of two neighboring peaks start to overlap, which roughly indicates the point where we can no longer reliably assign an atom number to an event.
} 
\end{figure}

Fig. \ref{Fig2}(b) shows a histogram of these corrected fluorescence levels, which we then fit with a sum of gaussians with centers $x_n$ and standard deviations $\sigma_n$.
We classify events within $x_n\pm 2\sigma_n$ to have $n$ atoms present; events outside these limits are discarded.
For the data shown in Fig. \ref{Fig2}, about 6\% of the events are rejected.
Tighter classification intervals lead to higher identification fidelity, but to a larger fraction of events being rejected.
We find that the mean fluorescence level $x_n$ grows linearly with the atom number $n$, as shown in Figure \ref{Fig3}(a), indicating no sign of density-dependent effects.

\section{Resolution limit and classification fidelity \label{section:fidelity}}

After establishing our post-processing procedure we now analyze the data to determine the fidelity of our atom number determination and find the limiting factors of the detection scheme. 
For our evaluation we assume that the peaks in the histogram are clearly distinguishable and we can assign reliable atom numbers as long as the separation of the peaks is larger than $4\sigma$.
Hence the maximal atom number we can resolve is limited by the observed growth of the peak width $\sigma_N$ with the atom number $N$, see Fig. \ref{Fig3}(b).

The peak width is determined by three noise sources, which can be assumed to be independent and therefore are added up quadratically \cite{Oberthaler2015double}:
These are a constant background noise stemming from the shot noise of the background light and the read-out noise of the camera, the shot noise of the fluorescence photons collected by the camera, which scales with the square root of the atom number, and shot-to-shot fluctuations in the average fluorescence rate per atom, which scale linearly with the atom number.

When analyzing the data we find that for atom numbers $<20$ the shot noise of the fluorescence photons contributes $\sigma_{ph}^2\lesssim 0.01^2$ to the variance and the atom-number independent background noise is about $\sigma_{cam}^2\approx 0.07^2$. 
These noise sources therefore only play a small role in in limiting our atom number determination.

Instead, the dominant effect is a linear growth of the peak width $\sigma_N$ with the atom number $N$, see Fig. \ref{Fig3}(b), which strongly suggests that our atom number determination is limited by fluctuations in the atom fluorescence.
While long-term drifts are removed by the rescaling method described in section \ref{section:evaluation}, shot-to-shot fluctuations or drifts on timescales shorter than a few tens of shots are not significantly suppressed with this method.
The observed level of fluctuations is compatible with laser frequency noise with an rms level of $\sim$150 kHz or rms beam power fluctuations of $\sim$3 \%.
We note that one-color schemes \cite{Kimble1994,Meschede1996,Grangier2001sub,Jochim2011deterministic,Oberthaler2013accurate,Oberthaler2015double,Klempt2019preparation} are less sensitive to frequency or power fluctuations due to the high saturation of the single cooling transition. 
In contrast, in our two color scheme the F = 9/2 ground state is addressed by two transitions, i.e. the D2 cooler and the D1 mF-pumper, which thus compete with each other. 
Hence, relative power fluctuations directly translate into fluctuations of the fluorescence per atom and lead to a faster growth of $\sigma_N$ than in one-color schemes.



For the data set shown in Fig. \ref{Fig3} the threshold value $\sigma_N = 0.25$ where the $\pm 2\sigma$ intervals of neighboring histogram peaks start to overlap is reached for $N\approx17$.
Beyond this threshold, we define the boundary between $N$ and $N+1$ atoms as $(x_N+2\sigma_N+x_{N+1}-2\sigma_{N+1})/2$, and our ability to separate the fluorescence signals belonging to different atom numbers substantially decreases.
We therefore consider $\sigma_N < 0.25$ to roughly indicate the threshold for fluorescence measurements with single atom number resolution.
We note that this condition is significantly more conservative than the requirement $\sigma_N < 1$ used by other authors \cite{Oberthaler2013accurate,Oberthaler2015double, Klempt2019preparation}. 
Extrapolating our observed peak width growth indicates that we reach $\sigma_N = 1$ for $N\approx 80$. 


\begin{figure}
    \center
    \includegraphics[width=8.6cm]{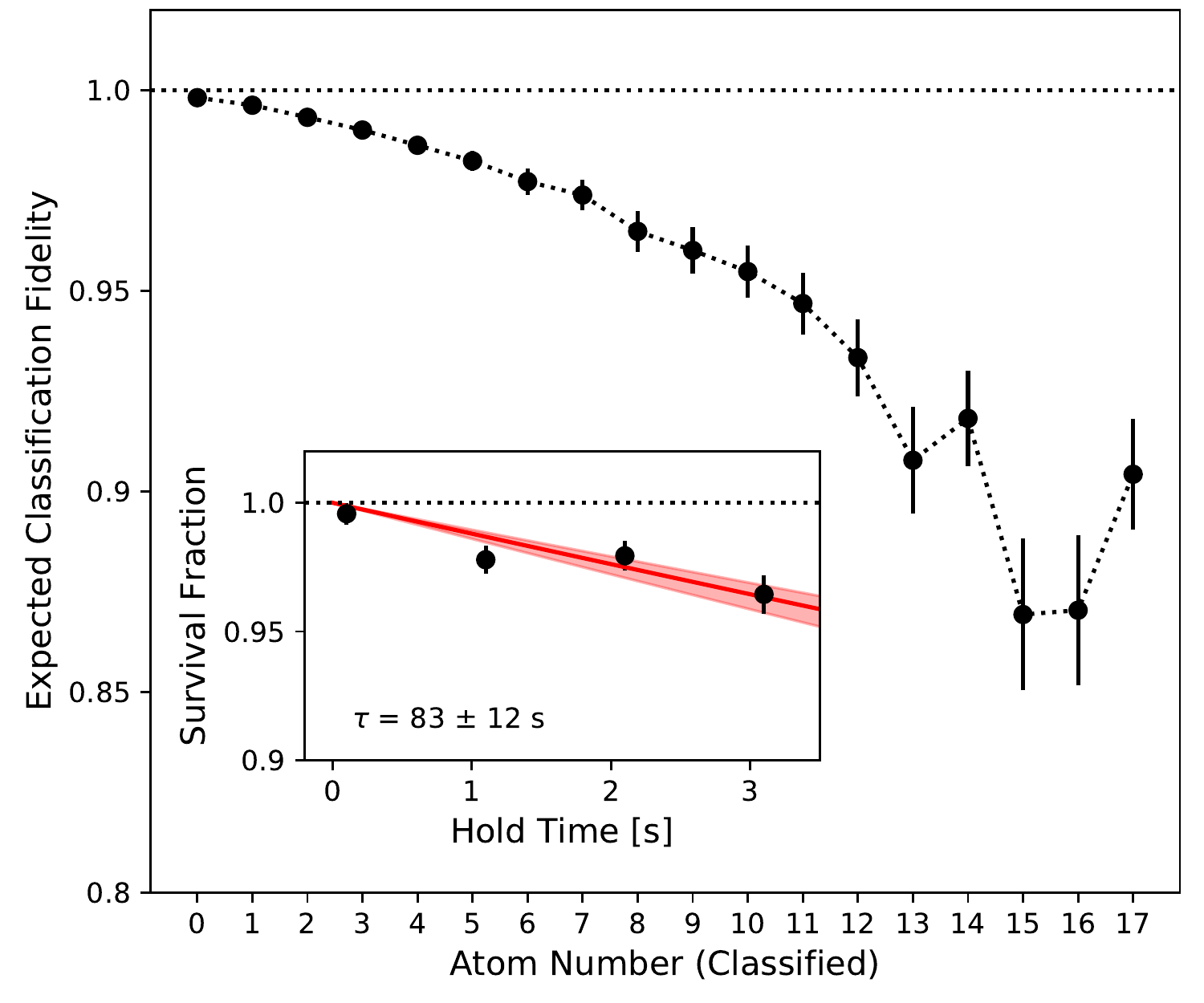}
    \caption{\label{Fig4} 
Simulated classification fidelities for systems with different atom numbers using the noise characteristics determined from in Fig. \ref{Fig2}(b) and an independent measurement of the MOT lifetime (see inset).
The classification fidelity is limited by the ratio of imaging time to life time: loss events during the imaging time reduce the number of collected fluorescence photons and thus can cause the fluorescence signal to fall into the classification bin of a lower atom number.
}
\end{figure}

In addition to the noise sources discussed above, which broaden the fluorescence peaks and therefore limit our ability to reliably distinguish between different atom numbers for larger values of $N$, there is another effect that can limit the fidelity of our atom number determination:
loss of atoms from the MOT during the imaging period.
As the lifetime of atoms in our imaging MOT is on the order of $80$\,s, much longer than the imaging time of ~1\,s, loss events are very rare when measuring small atom numbers, but become more relevant  for measuring larger systems.
Also, depending on the number of atoms in the system and at which time during the measurement it occurs, not every loss event necessarily leads to an incorrect determination of the atom number.
To obtain a quantitative estimate of the influence of the various noise sources on our classification fidelity we therefore perform a phenomenological Monte-Carlo simulation of our system using the following model for the counts $c_i$ oberserved in run $i$:

\begin{equation}
c_i = \alpha_i + \beta_i \cdot \sum_{n=1}^N \gamma_{i,n}\cdot \text{min}\left(1,\frac{\tau_{i,n}}{t_{Im}}\right),
\end{equation}
where $\alpha_i$, $\beta_i$, and $\gamma_{i,n}$ are gaussian distributed random numbers with mean 0, 1, and 1, respectively.
The standard deviations $\sigma_\alpha$, $\sigma_\beta$, and $\sigma_\gamma$ of these three random numbers are free parameters that model the noise sources of the fluorescence signal described above, where $\sigma_\alpha$ corresponds to the atom-number independent noise, $\sigma_\beta$ to the global and hence correlated fluctuations in fluorescence per atom, and $\sigma_\gamma$ to the fluorescence shot noise from each uncorrelated atom.
The values of these parameters are chosen such as to best reproduce the observed histograms \footnote{$\sigma_{\alpha} = 0.07 \pm 0.01$, $\sigma_{\beta} = 0.034 \pm 0.012$, $\sigma_\gamma = 0.012 \pm 0.005$}.
The possibility of atom loss is taken into account by comparing the imaging time $t_{Im}$ to a time $\tau_{i, n}$ after which atom $n$ is lost from the MOT in run $i$ that is drawn from an exponential distribution using the experimental lifetime of $83 \pm 12$ s, and taking the shorter of the two times to calculate the fluorescence of each atom \footnote{The lifetime has been measured by loading 0, 1, or 2 atoms into the MOT and comparing the fluorescence level of two measurements with some variable time delay between them, see inset of Fig. \ref{Fig4}. In the $\sim$4000 runs used for this, not a single event of loading an atom from the background gas has been observed, thus limiting the loading rate to be below $3\cdot10^{-4} s^{-1}$ at 68 \% CL.}. 
The counts from the simulation are processed and classified as described in section \ref{section:evaluation}.
The classification fidelity is estimated by comparing the initial atom number with the classification result for each simulated run.

Figure \ref{Fig4} shows the resulting classification fidelities, which are close to 100 \% for low atom numbers, decrease to $\sim$95 \% for 10 atoms and to $\sim$87 \% for the highest atom numbers we can reliably distinguish.
When working with at most 10 (17) atoms, the average fidelity is $\sim$98 \% ($\sim$95 \%).
The vast majority of the incorrectly classified events stem from atom loss, especially having one more atom initially than classified: when working with at most 10 (17) atoms, this is the case for $>99$ \% ($>95$ \%) of the incorrectly identified atom numbers. 

One interesting insight that can be gained from the simulation is that while the classification fidelity strongly depends on the MOT lifetime, this is not directly visible in the peak widths: a limited lifetime leads to a broad underground in simulated histograms emulating the one shown in Fig. \ref{Fig2}(b).
The peaks are still quite narrow and clearly identifiable, but loss of atoms during imaging can move the fluorescence count into the classification region of a lower atom number.
This can in principle be improved by increasing the lifetime of our MOT as the fraction of incorrectly identified events scales linearly with the inverse lifetime, but in our case the MOT lifetime is already close to the vacuum limited lifetime of the system.
Therefore, it is notable that reducing the width of the fluorescence peaks also improves the classification fidelity: smaller peak widths reduce the probability of misclassifying events involving atom loss and thus increase the fidelity.

\section{Conclusion\label{section:conclusion}}

In this work we have presented a novel two-color MOT for counting small numbers of atoms.
With this scheme, we are able to reliably perform atom number measurements with single particle resolution for systems of up to about 17 atoms with classification fidelities of on average 95\%. 
The number of atoms we can currently distinguish is determined by the stability of the fluorescence rate on the probing transition, and thus by the technical performance of the frequency locks and power stabilizations used, which can be improved if a need to count larger atom numbers arises.
The two-color scheme significantly reduces the influence of light scattered from the MOT beams and thereby enables such single-atom resolved measurements in systems that have not been purpose-built for this application and for example contain in-vacuum components.

\begin{acknowledgments}
We thank P. Wieburg for his work in building and characterizing the quantum gas apparatus.
This work is supported by the Deutsche Forschungsgemeinschaft (DFG, German Research Foundation) in the framework of SFB 925 and the excellence cluster 'Advanced Imaging of Matter' - EXC 2056 - project ID 390715994.
\end{acknowledgments}

\bibliography{IMOT}

\end{document}